\documentclass[12pt]{article}

\usepackage{a4wide}
\usepackage{amssymb}
\usepackage{slashed}
\usepackage{verbatim}

\usepackage[intlimits]{amsmath}

\usepackage{amsfonts}

\usepackage{epsfig}

\newcommand{\be}{\begin{equation}}
\newcommand{\ee}{\end{equation}}
\newcommand{\bea}{\begin{eqnarray}}
\newcommand{\eea}{\end{eqnarray}}
\newcommand{\ba}{\begin{eqnarray}}
\newcommand{\ea}{\end{eqnarray}}
\newcommand{\n}{\nonumber }

\begin{document}

\setcounter{table}{0}

\begin{flushright}\footnotesize

\texttt{ICCUB-20-004}

\end{flushright}

\mbox{}
\vspace{0truecm}
\linespread{1.1}

\vspace{0.5truecm}

\centerline{\Large \bf Displaced orbits and electric-magnetic black hole binaries}

\vspace{1.3truecm}

\centerline{
    {\large \bf Jorge G. Russo} }

\vspace{0.8cm}

\noindent  
\centerline {\it Instituci\'o Catalana de Recerca i Estudis Avan\c{c}ats (ICREA), }
\centerline{\it Pg. Lluis Companys, 23, 08010 Barcelona, Spain.}

\medskip
\noindent 
\centerline{\it  Departament de F\' \i sica Cu\' antica i Astrof\'\i sica and Institut de Ci\`encies del Cosmos,}
\centerline{\it Universitat de Barcelona, Mart\'i Franqu\`es, 1, 08028 Barcelona, Spain. }

\medskip

\centerline{  {\it E-Mail:}  {\texttt jorge.russo@icrea.cat} }

\vspace{1.2cm}

\centerline{\bf ABSTRACT}
\medskip

In presence of magnetic fields, the orbits of charged particles can be displaced from the equatorial plane.
We study circular orbits of electrically charged massive objects around a magnetic black hole in the probe approximation.
We show that there exist a one-parameter family of circular orbits at constant polar angle $\theta $ and constant radius $r$, parameterized by the angular momentum. The angle $\theta $ can be made arbitrarily small by increasing the charge-to-mass ratio of the orbiting particle, but when this is  
a Reissner-Nordstr\" om black hole, the condition $q\leq m$ implies that orbits exist only for $|\theta-\frac{\pi}{2}|<
 {\rm arccot}(2\sqrt{2})$. We show that the circular orbits are stable under small perturbations in the $\theta $ and $r$ directions.
We also discuss the Newtonian approximation and 
a binary system of electric and magnetic black holes, each one describing a circular orbit with no central force.

\noindent

\vskip 1.2cm
\newpage

\def\sech{ {\rm sech}}
\def\p{\partial}
\def\pa{\partial}
\def\ov{\over }
\def\a{\alpha }
\def\g{\gamma}
\def\s{\sigma }
\def\td{\tilde }
\def\vp{\varphi}
\def\strokedint{\int}
\def \ha {{1 \over 2}}

\def\KK{{\cal K}}




\textwidth = 450pt
\hoffset=-10pt







The study of the dynamics of test particles in black holes underlies a number of physically relevant phenomena, which
include the general classification of orbits, light deflection,  precession of periastro and Lense-Thirring effect.
The Reissner-Norstr\" om (RN) solutions arise as a solution of Einstein-Maxwell theory and can carry electric and magnetic charges.
Although they are not relevant in astrophysics, they have  been extensively investigated because of their multiple applications
in different areas of physics. In particular,  they play central role in the microscopic derivation of the black hole entropy, and
generalizations of RN such as charged black branes are fundamental in holographic dualities.

The geodesics of  general  spherically symmetric black holes can be found analytically in terms of elliptic and hyperelliptic functions.
The case of the Schwarzschild black hole was treated originally in 1931 by Hagihara [1], where the motion of test particles was described in
terms of  Weierstrass elliptic functions. But  important progress was made in the last decade, starting with the works  by
Hackman and L\"ammerzahl \cite{Hackmann:2008zza,Hackmann:2008zz}, where the method of the Jacobi inversion problem was used to analytically determine geodesics
of the Schwarzschild-de Sitter black hole in terms of hyperelliptic functions. The method was then 
extended to Taub-Nut and Kerr-de Sitter spaces \cite{Kagramanova:2010bk,Hackmann:2010zz}.

Trajectories of test particles in the RN spacetime have been studied extensively  (some recent works can be found in 
 {\it e.g.} \cite{Kim:2007ca,Pradhan:2010ws,Grunau:2010gd,Pugliese:2011py,Pugliese:2013xfa,Gonzalez:2017kxt}). In particular, Grunau and Kagramanova \cite{Grunau:2010gd}
studied the orbits of dyons, carrying  electric and magnetic charges $(q,p)$,
in the background of a dyonic RN black hole carrying both electric and magnetic charges $(Q,P)$.
Generic orbits were determined analytically in terms of elliptic  functions and a classification
of the different solutions was given. Trajectories are confined in cones, a feature that appeared in earlier studies of magnetic monopoles 
\cite{Golo:1982nu,LyndenBell:1996xj}. 
The case studied in this paper corresponds to the particular case $p=0$, $Q=0$; that is, 
  we shall consider the  motion of an electrically charged massive  particle
in the magnetic RN  background.  The system can also be viewed in terms of the electric-magnetic dual configuration, 
a magnetic monopole test particle in the background of an electric RN black hole.
Here we will focus on circular orbits at constant polar angle $\theta $ and constant radius $r$.
The circular orbits presented here do not appear anywhere in \cite{Grunau:2010gd} and, despite their simplicity, are new.
The circular orbits  were missed in the general analysis of 
\cite{Grunau:2010gd} because they correspond to a special case
where the right hand side of (26) in \cite{Grunau:2010gd} is identically zero.
As a result, the circular solution would correspond to  a singular limit where $\gamma'_{\rm in}\to \infty $. Understanding the circular solutions requires a careful analysis of the allowed
range of radii where circular orbits exist, the precise determination of the dependence of radii in terms of the conserved angular momentum and charges and finding the conditions for the stability of the orbits under small perturbations. This analysis will be carried out in this paper.
The circular solution exhibits some intuitive features regarding the interplay between the Lorentz force and the gravitational force
and have a number of interesting physical implications that we shall describe.

Our starting point is the magnetic RN solution of Einstein-Maxwell theory,
\be
ds^2 = -\lambda(r)\, dt^2 + \frac{dr^2}{\lambda(r)} +r^2 d\Omega_2^2\ ,\qquad 
A=P \cos\theta \, d\varphi \ ,
\ee
\be
\lambda(r) = \left( 1-\frac{r_+}{r}\right) \left(
1-\frac{r_-}{r}\right)\ ,\quad 
G\, M=\frac12 (r_++r_-)\ ,\quad G\, P^2=r_+r_-\ ,
\nonumber
\ee
where $r_\pm =GM\pm \sqrt{G^2M^2 -GP^2}$ and $\sqrt{G}M\geq P$.
Consider now the world-line action for a massive electrically charged particle in this background
\ba
S &=& - m\int d\tau \left(\sqrt{-g_{\mu\nu}(x) \dot x^\mu \dot x^\nu }- \frac{q}{m}\, A_\mu \dot x^\mu\right)
\n\\
&=& -m\int dt \left(\sqrt{\lambda  - \frac{\dot r^2}{\lambda} -r^2\dot\theta^2 -r^2\sin^2\theta \dot \varphi ^2   }- \frac{q}{m}\, P  \cos\theta\  \dot \varphi 
\right)\ .
\ea
where we have set $\tau =t$.
Trajectories are determined by solving the Euler-Lagrange equations,
which are given by
\ba
\label{eqr}
&&\frac{1}{\Delta} ( \lambda'+ \frac{\dot r^2 \lambda' }{\lambda^2}- 2r \dot\theta^2 -2r \sin^2\theta \dot \varphi ^2  )+ \frac{d}{dt }\left( \frac{2\dot r  }{\lambda\Delta }\right)=0\ ,
\\
\label{eqtheta}
&&  - \frac{1}{\Delta}\, r^2 \sin\theta \cos\theta \dot \varphi ^2+ \frac{q}{m}\,  P\sin\theta \dot\varphi 
+  \frac{d}{dt }\left(  \frac{r^2\dot \theta}{\Delta}   \right)=0\ ,
\\
&&  \frac{d}{dt}\left( \frac{1}{\Delta} r^2\sin^2\theta \dot \varphi +\frac{q}{m}\, P \cos\theta    \right)=0\ ,
\label{eqvarphi}
\ea
with
\be
\Delta\equiv \sqrt{\lambda  - \frac{\dot r^2}{\lambda} -r^2\dot\theta^2 -r^2\sin^2\theta \dot \varphi ^2   }\ .
\nonumber
\ee
The Hamiltonian is
\be
 H= \pi_r \dot r +\pi_\theta\dot\theta +\pi_\varphi \dot \varphi -{\cal L}\ ,\qquad
{\cal L}=-m \Delta +q P \cos\theta \dot \varphi \ ,
\nonumber
\ee
with
\be
\pi_r=\frac{m \dot r}{\lambda \Delta }\ ,\qquad \pi_\theta=\frac{mr^2\dot \theta}{\Delta }\ ,
\nonumber
\ee
\be
 J=\pi_\varphi=\frac{mr^2\sin^2\theta \dot \varphi}{\Delta }+ q P \cos\theta\ .
\label{fias}
\ee
This gives
\be
H=\frac{m\lambda}{\Delta} = E\ ,
\ee
where $E$ is a constant representing the energy of the particle measured by an observer at rest at infinity. 
We note that the angular momentum is non-vanishing when
$\dot\varphi =0$, which is a familiar feature in systems carrying both electric and magnetic charges.

We will now focus on solutions with constant $r$ and constant $\theta $.
Solving \eqref{fias} for $\dot \varphi $ in terms of $J$, one finds
\be
 r^2\sin^2\theta\ \dot \varphi ^2 =\frac{ \lambda (J-q P\cos\theta )^2}
{(J- q P\cos\theta )^2+m^2r^2\sin^2\theta}\ .
\label{fifi}
\ee
The equations  for $r$ and $\theta $ can be found  by substituting
the above solution for $\dot \varphi $ into \eqref{eqr}, \eqref{eqtheta}.
Equivalently, we can derive them from the effective potential
represented by the squared Hamiltonian.
Substituting \eqref{fifi} into $\Delta $, the squared Hamiltonian  takes the form
\be
V (r,\theta)=H^2=\frac{\lambda(r)}{ r^2\sin^2\theta} \left( \big(J- q P \cos\theta \big)^2+m^2 r^2\sin^2\theta \right)\ .
\label{veffe}
\ee
Differentiating with respect to $\theta $, we find
\be
\frac{\partial V }{\partial\theta} =
\frac{2\lambda}{ r^2\sin^3\theta } \big(J- q P \cos\theta \big)
 \big( q P - J\cos\theta \big)\ .
\label{qeq}
\ee
The  solution  $\cos\theta    = J/(qP)$ 
is inconsistent with the $r$ equation as it leads to $\lambda'=0$, which does not have  a solution outside the horizon.
The unique solution with $\dot\varphi\neq 0$ is
\be
\cos\theta    = \frac{ q P}{J}\ .
\label{qqt}
\ee
%
By the symmetry of the configuration under $\theta\to \pi-\theta $, we may assume $0<\theta\leq \frac{\pi}{2}$ and $J>0$.
From \eqref{qqt} we see that  a necessary condition for circular orbits to exist is $J\geq q P$.
As shown below, this condition is not sufficient. 

Note that, for $P=0$, \eqref{qqt} gives $\theta =\pi/2$. Substituting $P=0$, $\theta =\pi/2$  into \eqref{veffe}, one gets the
familiar effective potential for the case of the Schwarzschild black hole \cite{Wald:1984rg}.

\begin{figure}[h!]
\centering
\includegraphics[width=0.5\textwidth]{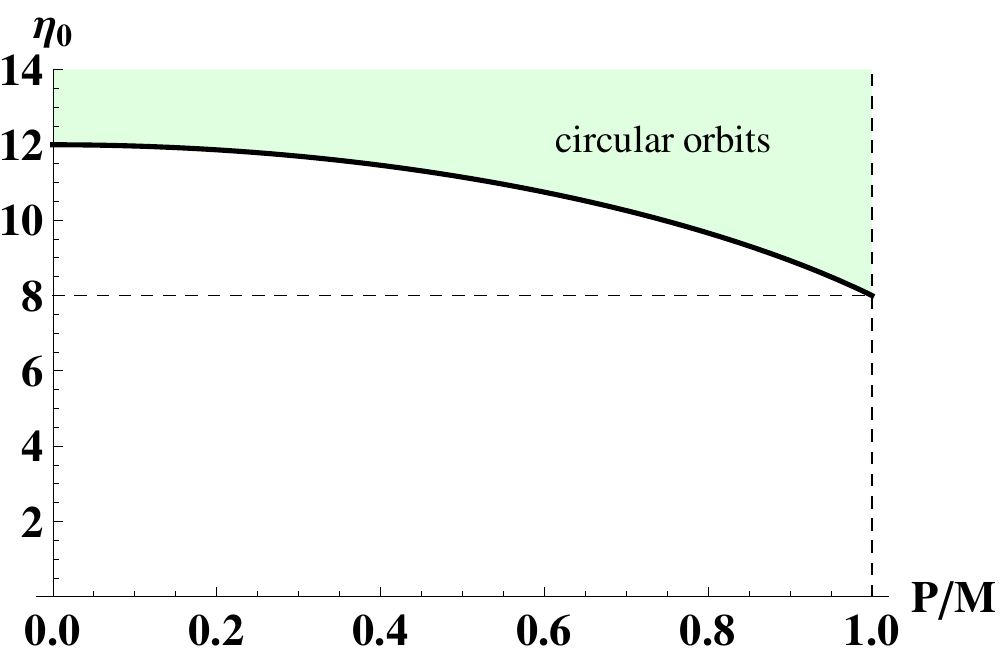}
\caption{The shaded area shows the region where circular orbits at constant $\theta $ exist. The thick line
 shows the value $\eta_0 (P/M)$ at which the orbits at $r_2$ and at $r_3$ coincide. This orbit represents the ISCO for a given $P/M$. }
\label{aapp}
\end{figure}

Substituting \eqref{qqt} into \eqref{veffe}, the effective potential takes the form
\be
V (r)=m^2 \lambda(r) \left( 1+\frac{A}{m^2r^2}\right)\ ,
 \label{veffer}
\ee
where $A\equiv J^2-q^2P^2$.
The equation for the radial variable is ($G=1$)
\be
\frac{\partial V }{\partial r}=0\ \  \longrightarrow
\ \ m^2 M r^3 -(A+m^2P^2) r^2+3AM r
-2 AP^2=0\ .
\label{radios}
\ee
Equations \eqref{qeq}, \eqref{radios} are equivalent to the Euler-Lagrange equations \eqref{eqr}, \eqref{eqtheta}.
Two orbits coincide when the  discriminant of the radial  equation vanishes. This gives the condition
\be
\eta ^3 \left(8 p^2-9\right)+\eta ^2 \left(24
   p^4-126 p^2+108\right)+\eta  \left(24 p^6-9 p^4\right)+8 p^8=0\ ,
\label{discri}
\ee
where $\eta \equiv A/(GmM)^2$ and $p\equiv P/M$.
When $0< P<M$, there are three real solutions $r_1< r_2< r_3$ provided  $\eta > \eta_0(P/M)$,  with $8<\eta_0(P/M)\leq 12$.
 $\eta_0(P/M)$
represents the root of  \eqref{discri} at which $r_2=r_3$ (see Fig.   \ref{aapp}).

\begin{figure}[h!]
\centering
\includegraphics[width=0.5\textwidth]{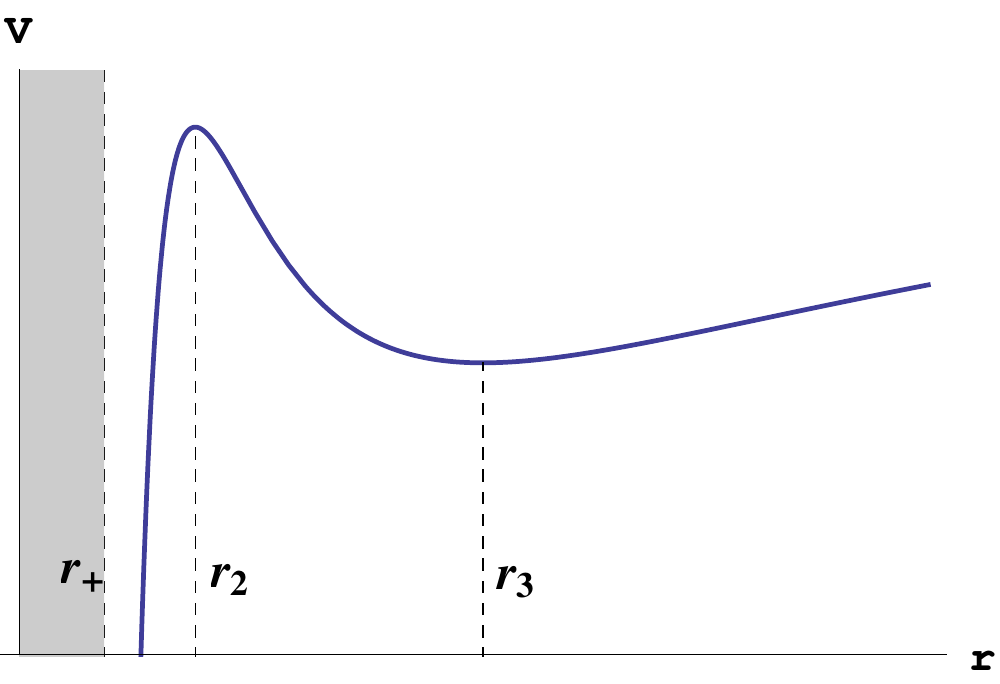}
\caption{Effective potential in the case $\eta =14$, $P=M/2$, exhibiting a maximum at $r=r_2$ and
a minimum at $r=r_3$. There is another minimum at $r=r_1$ lying inside the horizon.}
\label{vife}
\end{figure}

For given charges $P,\ q$ and masses $M,\ m$, the orbits are fully characterized
by the angular momentum $J$. Both $\theta $ and the radius $r$ of the stable orbit are uniquely given in terms of $J$.
In terms of the parameter $\eta $, the cases are as follows:

\begin{itemize}

\item For $P=0$ (Schwarzschild case), there are three real roots $r_1< r_2< r_3$ provided $\eta \to J^2/(GmM)^2> 12$.
$r_1=0$ and it does not represent an orbit.
$r_3$ corresponds to a minimum of the potential
and $r_2$ to a maximum.
Thus $r_3$ is the   only relevant stable circular orbit. At $\eta =12$, one has 
$r_2=r_3=6M$, which becomes an inflexion point of the potential. 
 For $J/(mM)\gg 1$, $r_3\approx J^2/(Mm^2)$ and $r_2\approx 3M$.

\item As mentioned above, when $0< P<M$, there are three real solutions $r_1< r_2< r_3$ as long as  $\eta > \eta_0(P/M)$.
For $\eta > \eta_0(P/M)$, the orbit at $r=r_3 $ is a minimum of the potential
and represents  the unique stable circular stable orbit. 
$r_1$ lies inside the horizon, $r_1< r_+$.
See Fig. \ref{vife}.

\item At  $\eta =\eta_0(P/M)$,   $r_2$, $r_3$ coincide, 
representing an inflexion point of the potential.
Therefore the orbit  is marginally stable (unstable under second-order
perturbations).

\item In the extremal limit $P\to M$,  $\eta_0\to 8$.
For $\eta> 8$, there is a stable circular orbit at $r_3$.
As $\eta\to 8$, $r_2,r_3\to 4M$.
 
\item If $\eta <8$, there are no orbits for any $P$ in the interval $0\leq P\leq M$. The roots $r_2$ and $r_3$ are imaginary. 

\end{itemize}

\begin{figure}[h!]
\centering
\includegraphics[width=0.5\textwidth]{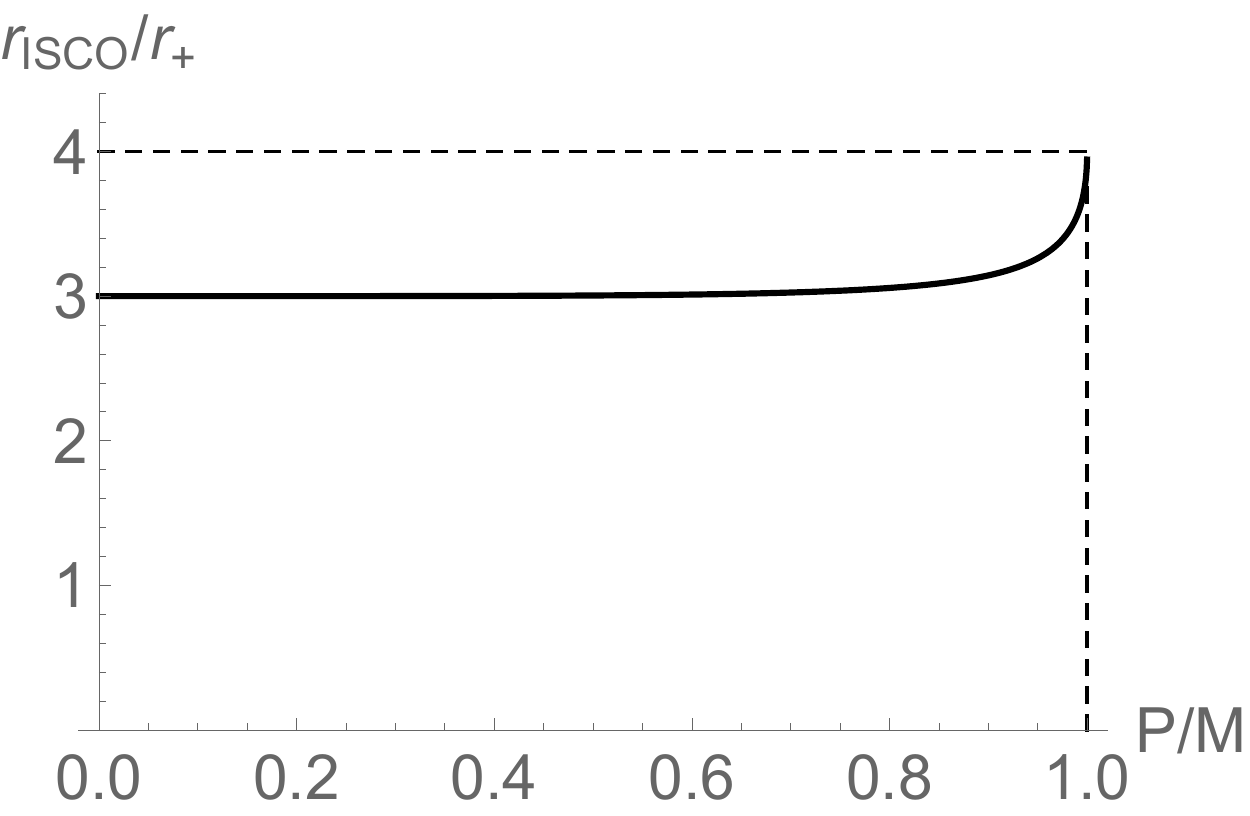}
\caption{The ratio $r/r_+$ for the innermost stable circular orbit as a function of $P/M$. }
\label{iscor}
\end{figure}

Thus 
stable circular orbits exist in the shaded region of figure \ref{aapp}. In terms of the angular momentum, this
 gives the condition $J>\sqrt{q^2P^2+ (mM)^2\eta_0(P/M)}$,
with $8<\eta_0(P/M)\leq 12$.
The innermost stable circular orbit (ISCO) depends on the value of $P/M$. It lies at a radius $r=r_3$, which
varies monotonically between $6M$ and $4M$ as $P/M$ is increased from 0 to 1 (see Fig. \ref{iscor}).

\begin{figure}[h!]
\centering
\includegraphics[width=0.24\textwidth]{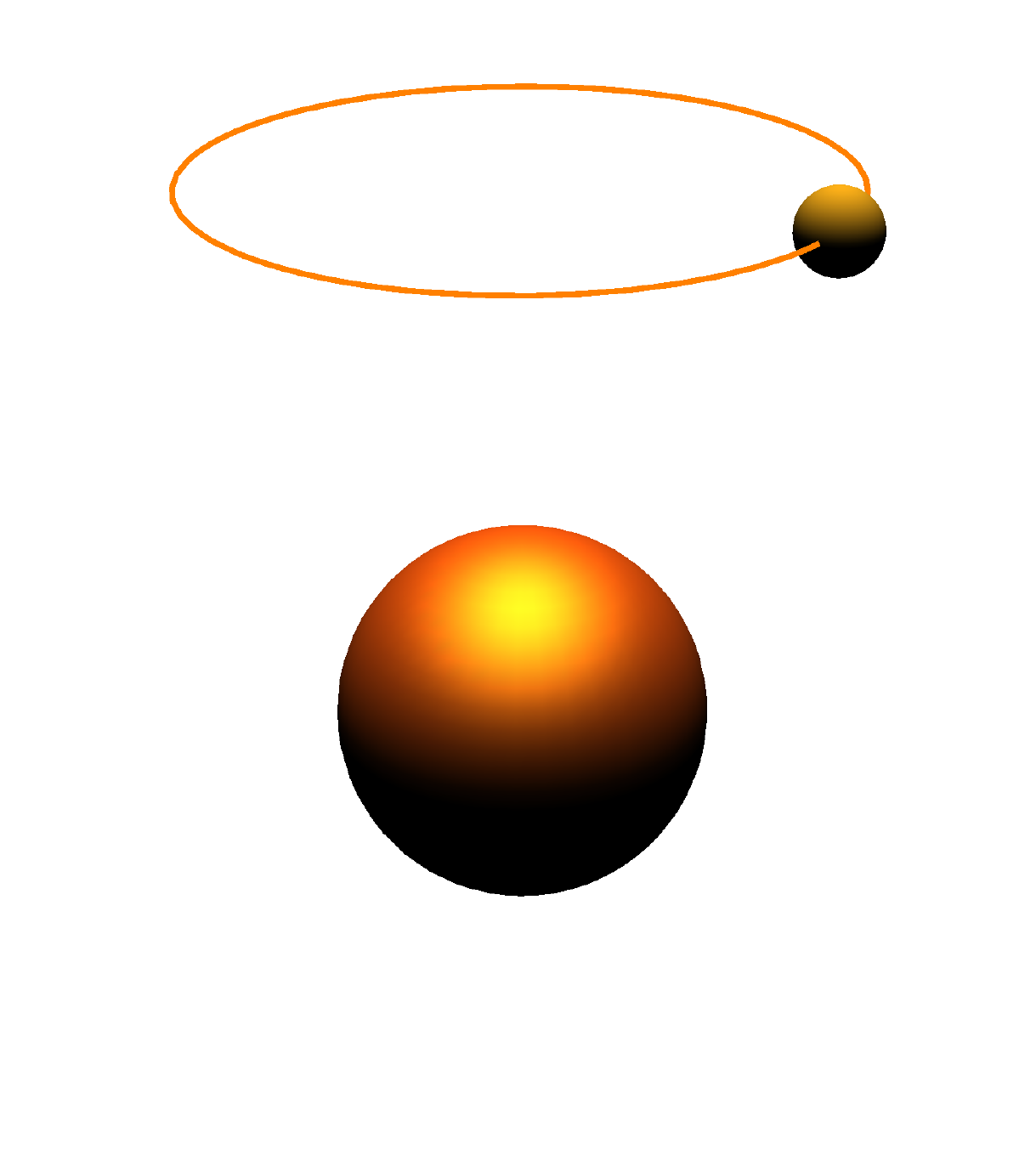}
\caption{Exact scale (1:1) plot of the
orbit  of an electrically ($q=e$) charged massive particle around a magnetic black hole of  Earth mass $M$  and $P=0.9 \sqrt{G}M$ ($r_+\approx 0.64\, {\rm cm}$).
 Here $\eta =9$ and $m=0.32\, \mu {\rm g}$, giving an  orbit at  $r\approx 2.19\, {\rm cm}$ and
$\theta=\pi/6$.
}
\label{coronita}
\end{figure}

The  motion takes place along a circular orbit displaced from the equatorial plane 
of the magnetic black hole (see Fig. \ref{coronita}).
The Lorentz force is orthogonal to the radial direction and to the
tangent vector to the orbit. It has a component in the $z$-direction
that balances the $z$-component of the attractive gravitational force,
to give a net centripetal force pointing to the center of the circle.
The relation $\cos\theta =q P/J$ shows that the polar angle of the orbit $\theta $ becomes closer to 0 as the product $qP$, which controls the strength of 
 the Lorentz force,  increases for a given $J$.

Under electromagnetic duality, the configuration turns into a massive monopole orbiting an electric Reissner-Nordstr\" om black hole.
Note that $A$  and $\theta $ are manifestly invariant 
 under the exchange $q\leftrightarrow P$, 
which shows that the radius $r$ and constant polar angle $\theta $ of the orbit remain the same.

\medskip

Let us now consider stability under small perturbations in the $r$ and $\theta $ directions. Stability requires that the matrix of second derivatives of the potential, evaluated on the solution $\{ r= r_3, \ 
 \cos\theta= q P/J\} $, has positive eigenvalues.
Computing the second derivatives on the solution, we find
\be
\frac{\partial^2 V }{\partial r\partial\theta} =0\ .
\ee
As the matrix of second derivatives is diagonal, we only need to check positivity of 
$\frac{\partial^2 V }{\partial r^2}$ and $\frac{\partial^2 V }{\partial \theta^2}$. We find
\bea
&&\frac{\partial^2 V }{\partial \theta^2}= \frac{2J^2}{r_3^2}\,  \lambda(r_3) >0\ ,
\nonumber\\
&& \frac{\partial^2 V }{\partial r^2}=\frac{2}{r_3^6}\left( A\big( 6P^2+r_3(r_3-6M)\big) +P^2 m^2 r_3^2\right)\ ,
\label{Vrr}
\eea
The second derivative in the $\theta $ direction is positive definite outside
the horizon where the orbits at $r_3$ lies. In the second derivative with respect to $r$, we have used
the $r$ equation \eqref{radios} to eliminate $r^3$ in the numerator. The resulting expression \eqref{Vrr} is manifestly positive for $r_3 \geq 6M$, which is the case when  $\eta\geq 12$ for any $0< P\leq M$.
For $P=0$ and $r_3=6M$, which is achieved at $\eta=12$, 
$\partial^2 V/\partial r^2 $ vanishes and $r=r_3$
becomes an inflexion point, as discussed above.
In the interval $8<\eta <12$, positivity can be easily checked numerically
  in the shaded area of figure \ref{aapp}. The fact that $r=r_3$ is a minimum of the potential
can also be understood analytically in a simple way. Near infinity, the effective potential is $V \approx 1-2M/r +O(1/r^2)$,
so $V $ decreases as $r$ is decreased from $\infty$. Therefore the first extremum coming from infinity, 
corresponding to $r_3$, must be a minimum. 

Small perturbations in $\theta $ and in $r$ from their equilibrium values will
lead to oscillatory motion in these directions with squared frequencies proportional
to the $\frac{\partial^2 V }{\partial r^2}$ and $\frac{\partial^2 V }{\partial \theta^2}$ computed above. Since these frequencies are different
from $\dot \varphi $, this will lead to a precession rate of the radial coordinate and a Lense-Thirring precession in the polar angle  $\theta $.

Let us now discuss the possible values of $\theta $.
For an electron,  $q/(m\sqrt{G})$ is very large; a computation gives $q/(m\sqrt{4\pi\epsilon_0 G}) \approx 2.04\times 10^{21}$.
As a result, the angle $\theta $ can be tiny, electrons can have orbits in  small circles. This is just a reflection of the fact  
that the electromagnetic force is much stronger than the gravitational force for electrons.
The situation is very different  when the orbiting particle is a Reissner-Nordstr\" om black hole
with $q\leq m\sqrt{G}$. Setting $G=1$, in
 terms of $\theta $, we have
$\eta=\frac{ q^2P^2}{m^2M^2} \tan^2\theta\ .$
The conditions $\eta>8$, $q\leq m$, $P\leq M$ then imply a minimum angle
$\theta_0 $ for the existence of the orbit, with
\be
\tan^2\theta_0 = 8\ \ \longrightarrow\ \ \theta_0={\rm arctan}(2\sqrt{2}) \ ,
\ee
and possible orbits lie in the interval $|\frac{\pi}{2}-\theta |<
 {\rm arccot}(2\sqrt{2})\approx 0.108 \pi$.

\medskip


The magnetic Reissner-Nordst\" om spacetime may be viewed as
a  ``cosmic" mass spectrometer, by virtue of the property that, 
in the presence of a magnetic field,  trajectories of particles depend on the ratio $q/m$.
As trajectories of different particles also depend on initial conditions, an efficient design requires fixing some
constants of motion to be the same for all particles.
Consider  two particles 1 and 2 
in the magnetic RN background, orbiting at the {\it same} radial distance $r$. We would like to determine
how the polar angle depends on $q/m$ of each particle.
Solving the radial equation \eqref{radios} for $A$ leads to
\be
\eta =\frac{q^2P^2}{m^2M^2}\tan^2\theta=\frac{r^2(Mr-P^2)}{M^2(r^2-3Mr+2P^2)}\ .
\ee
Therefore all particles orbiting at the same $r$ have the same parameter $\eta $, irrespective of the value of  $q/m$.
This implies the simple relation:
\be
\frac{\tan\theta_1}{\tan\theta_2}  =\frac{ m_1/q_1}{m_2/q_2}\ .
\ee
Here we omitted absolute value bars. Thus, by measuring the angle of such equal-radius orbits, one  determines the  ratios of $q/m$.
If the particles orbit at different radial distances, one needs more information to determine  $q/m$.
In  laboratory spectrometers, ions are accelerated at the same kinetic energy. In this spirit, one could here
fix $E^2$ for both particles; in this case  they  orbit at different $r$ and
the ratios of $q/m$ are given by a more complicated formula. Some natural ``cosmic" mass spectrometers  are provided by astrophysical black holes
embedded in magnetic fields  or by the magnetosphere of neutron stars, where the magnetic fields can significantly affect the trajectory of ionized matter
 \cite{Hackstein:2019msh}.

\medskip

\begin{figure}[h!]
\centering
\includegraphics[width=0.5\textwidth]{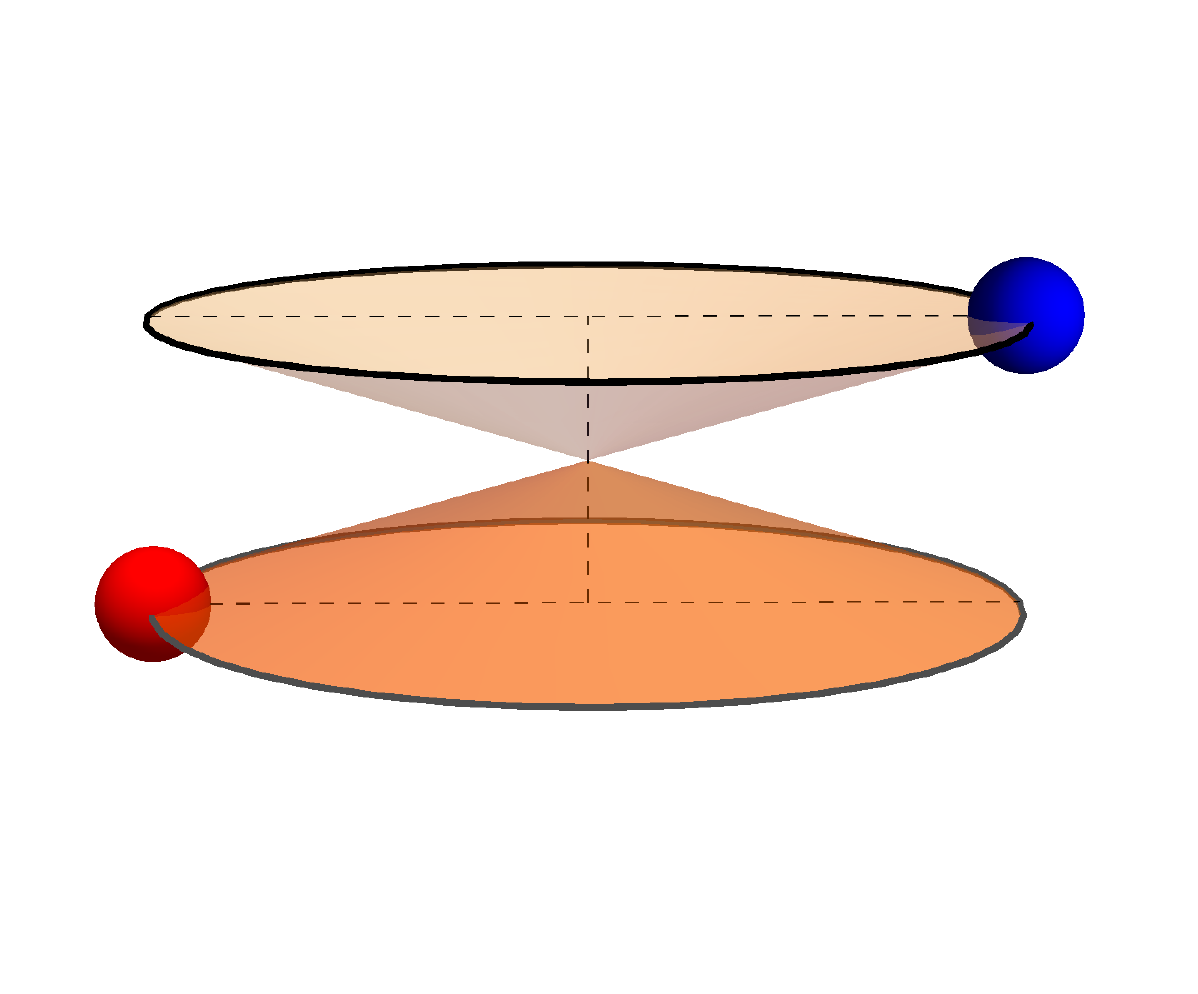}
\caption{Binary system for a pair of electric and magnetic compact objects of equal masses.
The orbits describe circular motion with  ``nothing" in the center.
 }
\label{binario}
\end{figure}

\medskip
Going beyond the probe approximation is obviously complicated, because it requires solving Einstein equations analytically
for two orbiting massive objects. 
However,  at large distances, velocities are non-relativistic and the orbits can be computed by a Newtonian analysis of forces.
Consider  an electric-magnetic binary system consisting  of a magnetically  charged particle and an electrically  charged particle
 of 
 equal masses $M$. 
The combination of Lorentz   and gravitational forces leads to the orbit of figure \ref{binario}. 
The analysis of equilibrium of forces is the same in each particle,
because the Lorentz force depends on the product $PQ$ which is invariant under the exchange of electric and magnetic charges. Therefore we only need to focus on one of the particles.
The Lorentz force acting on the electric particle is
\be
 \vec F_{\rm Lorentz} =\frac{Q}{c}\, \vec v\times \vec B,\qquad \vec B= P \frac{\vec r}{r^3}\ ,\quad 
\vec v=\hat\varphi \, \dot \varphi\, \frac{r}{2}\sin\theta \ .
\ee
  We keep the order $|v|/c$ in the Lorentz force but neglect $O(v^2/c^2)$ terms. To this order,
the gravitational interaction is given by the Newton force $\vec F_{\rm grav} = -GM^2 \vec r/r^3$ (the contribution of the magnetic energy is negligible
at large distances because it decreases as $P^2/r$). The sum of the two forces must match the centripetal force $=M\dot\varphi^2 (r/2)\sin\theta $, which has
a component only in the radial direction of cylindrical coordinates.
This leads to the following formulas for 
$\theta $ and the separation distance $r$ in terms of the total angular momentum $J$, charges and masses:
\be
\cos\theta \approx \frac{QP}{2J}\ ,\quad r\approx \frac{4J^2}{GM^3}\ ,\quad J=2M\dot\varphi (\frac12 r\sin\theta )^2\ ,\quad  J\gg GM^2\ ,
\ee
The resulting configuration is illustrated in figure 
\ref{binario}.
The motion is non-relativistic,
as the velocity is $|v|=O(GM^2/J)\ll 1$. Consistency of the approximation requires $r\gg r_+$, which  is automatically implied by $J\gg GM^2$.

Note that the gravitational force decreases as $GM^2/r^2$, whereas the Lorentz force has a faster decrease as $O(1/r^{\frac52})$
(the magnetic field decreases as $P/r^2$ and the velocity decreases as $1/r^{\frac12}$). Therefore, at shorter distances,
the effect of the Lorentz force is stronger and can thus  support  orbits with smaller $\theta $, i.e. further away from the equatorial plane.


The electric-magnetic binary system of  black holes of equal masses can  be presumably studied 
numerically. This may shed light  on the behavior at shorter distances and on the conditions for the
stability of the orbit.

Similar constant radius orbits as those  studied here should exist in the presence of a cosmological constant, that is, 
for a charged particle moving in the background of a de Sitter or anti de Sitter magnetic black hole.
We expect many interesting effects to occur in these cases, since the radial equation involves a higher degree polynomial.

\subsection*{Acknowledgments}
%
The author is grateful to Paul Townsend for useful comments.
 He would like to  especially thank Armun Liaghat for many valuable discussions. We acknowledge financial support from projects 2017-SGR-929, MINECO
grant FPA2016-76005-C.

\end{document}